\documentclass[12pt,preprint]{aastex}
\begin{document}

\title{High-Resolution Spectroscopy of Gamma-Ray Lines from the
X-Class Solar Flare of 23 July, 2002}

\author{D. M. Smith\altaffilmark{1}, G. H. Share\altaffilmark{2},
R. J. Murphy\altaffilmark{2},
R. A. Schwartz\altaffilmark{3,4}, A. Y. Shih\altaffilmark{1,5},
R. P. Lin\altaffilmark{1,5}}

\altaffiltext{1}{Space Sciences Laboratory, University of California, Berkeley, 
Berkeley, CA 94720}
\altaffiltext{2}{E. O. Hulbert Center for Space Research, Naval Research
Laboratory, Washington, DC 20375}
\altaffiltext{3}{NASA Goddard Space Flight Center, Greenbelt, MD, 20771}
\altaffiltext{4}{Science Systems \& Applications, Inc.}
\altaffiltext{5}{Department of Physics, University of California, Berkeley, 
Berkeley, CA 94720}

\begin{abstract}

The \it Reuven Ramaty High Energy Solar Spectroscopy Imager (RHESSI)
\rm has obtained the first high-resolution measurements of nuclear
de-excitation lines produced by energetic ions accelerated in a solar
flare, a GOES X4.8 event occurring on 23 July, 2002 at a heliocentric
angle of $\sim73^{\rm{o}}$.  Lines of neon, magnesium, silicon, iron,
carbon, and oxygen were resolved for the first time.  They exhibit
Doppler redshifts of 0.1--0.8\% and broadening of 0.1--2.1\% (FWHM),
generally decreasing with mass.  The measured redshifts are larger
than expected for a model of an interacting ion distribution isotropic
in the downward hemisphere in a radial magnetic field.  Possible
interpretations of the large redshifts include 1) an inclination of
the loop magnetic field to the solar surface so that the ion
distribution is oriented more directly away from the observer, and 2)
extreme beaming of the ions downward along a magnetic field normal to
the solar surface.  Bulk downward motion of the plasma in which the
accelerated ions interact can be ruled out.

\end{abstract}

\keywords{Sun:flares --- Sun:X-rays, gamma rays --- line:profiles --- 
gamma rays:observations}

\section{Introduction}

We report the first high-energy-resolution measurements of nuclear
de-excitation lines in a solar flare.  These lines were detected by the \it
Reuven Ramaty High Energy Solar Spectroscopic Imager (RHESSI) \rm during
the GOES X4.8-class solar flare of 23 July, 2002, which
occurred at coordinates S13E72 on the solar disk
($73^{\rm{o}}$ heliocentric angle).  The characteristics of
this flare are reviewed in depth in \citet{Li03}.

Gamma-ray lines from inelastic interactions of accelerated ions with
ambient nuclei in solar flares \citep{Li67} were first observed by
\citet{Ch73}.  The de-excitation lines can be Doppler shifted due to
the nuclear recoils from the ion interaction and the emission of the
gamma ray.  The line profiles thus reveal the angular distribution of
the interacting ions \citep{Ra76}.  Accelerated protons and
$\alpha$-particles produce narrow lines while accelerated heavy ions
produce broad lines which merge together to form an underlying nuclear
continuum.  \citet{Sh02} analyzed an ensemble of gamma-ray line flares
from the \it Solar Maximum Mission \rm Gamma-Ray Spectrometer (\it
SMM/\rm GRS).  They found that narrow lines exhibit $\sim$~1\%
redshift for flares at small heliocentric angles, but are not
appreciably shifted near the limb of the Sun.  Their measurements are
consistent with a distribution of interacting particles which is
isotropic in the downward hemisphere and zero in the upward hemisphere.

Although the \it SMM/\rm GRS data yielded useful measurements of some
nuclear-line redshifts and put constraints on the line widths, the
energy resolution of that instrument was moderate, averaging around
4\% FWHM over the nuclear-line range. \it RHESSI \rm \citep{Li02} was designed to
provide high-resolution measurements of the shape and redshift of
de-excitation lines using cryogenically cooled
germanium detectors to achieve an energy resolution averaging around
0.2\% FWHM from 1--6~MeV \citep{Sm02}.

\section{Analysis and Results}

The \it RHESSI \rm spectrometer consists of nine coaxial germanium
detectors, each divided into a $\sim$1~cm upper segment (which stops
most hard x-rays) and a $\sim$7~cm rear segment.  Here we use only
data from the rear segments, which have most of the efficiency for
stopping gamma-rays above 300~keV \citep{Sm02}.  Detector \#2, which
is operated in an unsegmented mode and has degraded energy resolution,
was not used.  The spectra in Figure~1 were accumulated from 00:27:20
to 00:43:20 UT on 23 July, 2002, an interval that includes most of the
high-energy emission from this flare \citep{Li03}.

The flare intensity is roughly half the background level above
$\sim$1~MeV.  The instrumental gamma-ray background is caused by the
interactions of cosmic rays and protons trapped in the radiation belts
with materials in the spacecraft and the atmosphere.  Every 15 orbits
(about one day), the geomagnetic coordinates of the spacecraft
approximately repeat themselves, and therefore so do the cosmic and
trapped particle fluxes.  Thus we model the background during the
flare with the average of two spectra, taken 15 orbits before and
after the flare, each of which is the same length as the flare
observation itself.  Above 3~MeV, however, we replaced this background
spectrum with a different one using a full day's integrated data.  In
this high-energy band the background is simple in form and nearly free
of lines, so that we can can get an accurate background estimate
with small statistical fluctuations, regardless of the exact orbital
parameters, simply by rescaling the full-day integration to match the
intensity of the $\pm$ 15-orbit spectrum.

\begin{deluxetable}{lllllllll}
\tabletypesize{\scriptsize}
\tablecolumns{7} 
\tablewidth{0pc} 
\tablecaption{Best fit Gaussian parameters for prompt nuclear lines}
\tablehead{ 
\colhead{Isotope} & \colhead{Rest Energy} & 
\colhead{Fit Energy} & \colhead{\% Redshift} & 
\colhead{FWHM} & \colhead{\% FWHM} & 
\colhead{Fluence} & \colhead{\it SMM/\rm GRS} & \colhead{\it SMM/\rm GRS} \\
\colhead{} & \colhead{(keV)} & 
\colhead{(keV)} & \colhead{} & 
\colhead{(keV)} & \colhead{} & 
\colhead{(ph~cm$^{-2}$)} & \colhead{\% Redshift\tablenotemark{a}} & \colhead{\% FWHM\tablenotemark{a}} }
\startdata
$^{56}$Fe &  847 & $ 846.09 ^{+0.70}_{-0.60}$ & $0.11 ^{+0.08}_{-0.07}$ & $  1.2 ^{+2.9}_{-1.1}$ & $0.14 ^{+0.34}_{-0.13}$ & $ 7.5 ^{+3.4}_{-2.3}$ & ... & ...\\  \\
$^{24}$Mg & 1369 & $1363.6 ^{+2.3}_{-2.0}$ & $0.40 ^{+0.17}_{-0.14}$ & $ 21.0 ^{+8.0}_{-5.4}$ & $1.54 ^{+0.59}_{-0.39}$ & $28.3 ^{+7.2}_{-6.6}$   & ... & ... \\ \\
$^{20}$Ne & 1634 & $1628.8 ^{+1.7}_{-1.7}$ & $0.32 ^{+0.10}_{-0.10}$ & $ 17.6 ^{+4.3}_{-3.6}$ & $1.07 ^{+0.26}_{-0.22}$ & $21.4 ^{+3.8}_{-4.5}$  & $0.0 \pm 0.2$ & $2.9\pm 1.0$  \\\\
$^{28}$Si & 1779 & $1776.8 ^{+1.9}_{-2.1}$ & $0.12 ^{+0.11}_{-0.12}$ & $ 16.7 ^{+4.5}_{-5.4}$ & $0.94 ^{+0.25}_{-0.30}$ & $17.1 ^{+4.0}_{-4.5}$ & ... & ... \\  \\
$^{12}$C & 4438 & $4403 ^{+10}_{-10}$ & $0.79 ^{+0.23}_{-0.22}$ & $ 92 ^{+42}_{-29}$ & $2.06 ^{+0.95}_{-0.65}$ & $28.6 ^{+13.1}_{-8.6}$ & $-0.07\pm 0.14$\tablenotemark{b} & $3.18\pm 0.74$ \\ \\
$^{16}$O & 6129 & $6094 ^{+15}_{-18}$ & $0.58 ^{+0.24}_{-0.29}$ & $122 ^{+68}_{-51}$ & $1.99 ^{+1.11}_{-0.83}$ & $34.2 ^{+12.8}_{-15.5}$ & $-0.26\pm 0.13$\tablenotemark{b} & ... \\
\enddata 
\tablenotetext{a}{Average of five flares near heliocentric angle 74$^{\rm{o}}$}
\tablenotetext{b}{Blueshift}
\end{deluxetable}

\begin{figure}
\plotone{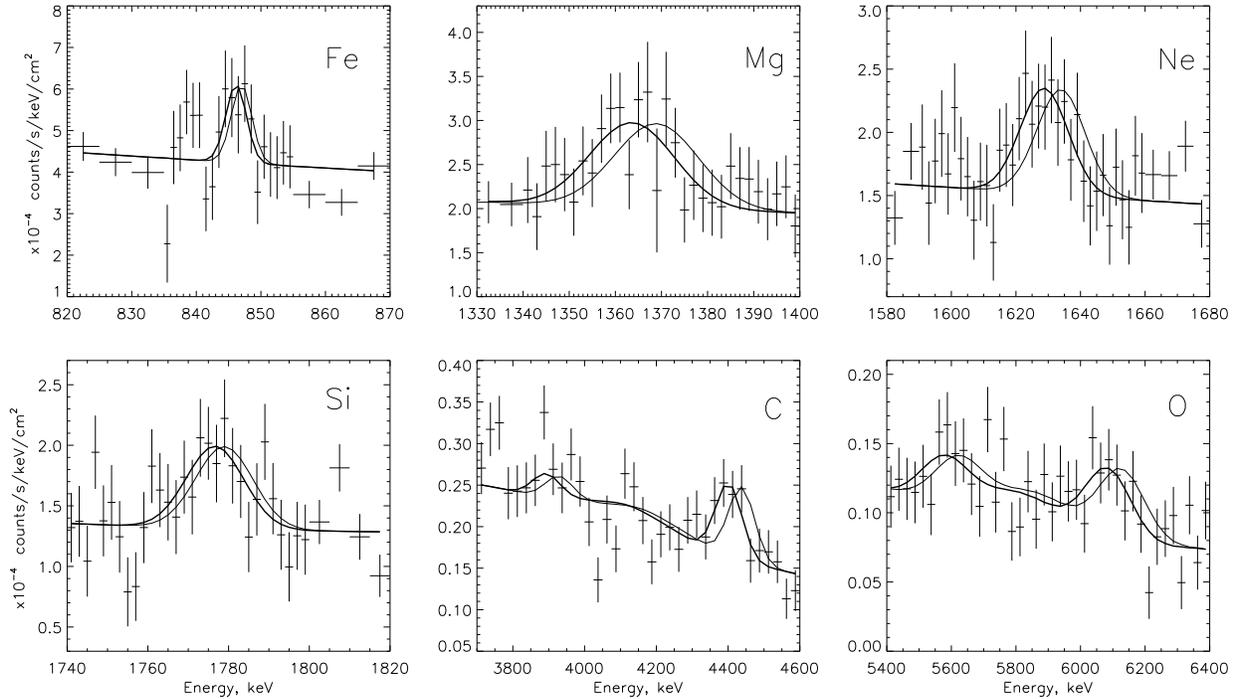}
\caption{\it RHESSI \rm background-subtracted count spectra
from 00:27:20 UT to 00:43:20 UT on 23 July, 2002.  Each panel
is labeled with the element primarily responsible for the line
shown.  The carbon and oxygen lines also show the secondary peak
from escape of a 511~keV positron-annihilation photon,
which also contains information on the line shape.  
The heavy curve shown in each panel is the Gaussian fit from Table~1
plus the underlying bremsstrahlung continuum and broad lines (see text),
convolved with the instrument response.
The lighter line is the same fit forced to zero redshift for
comparison.  The error bars are $\pm 1 \sigma$ from Poisson statistics.}
\end{figure}

\begin{figure}
\plotone{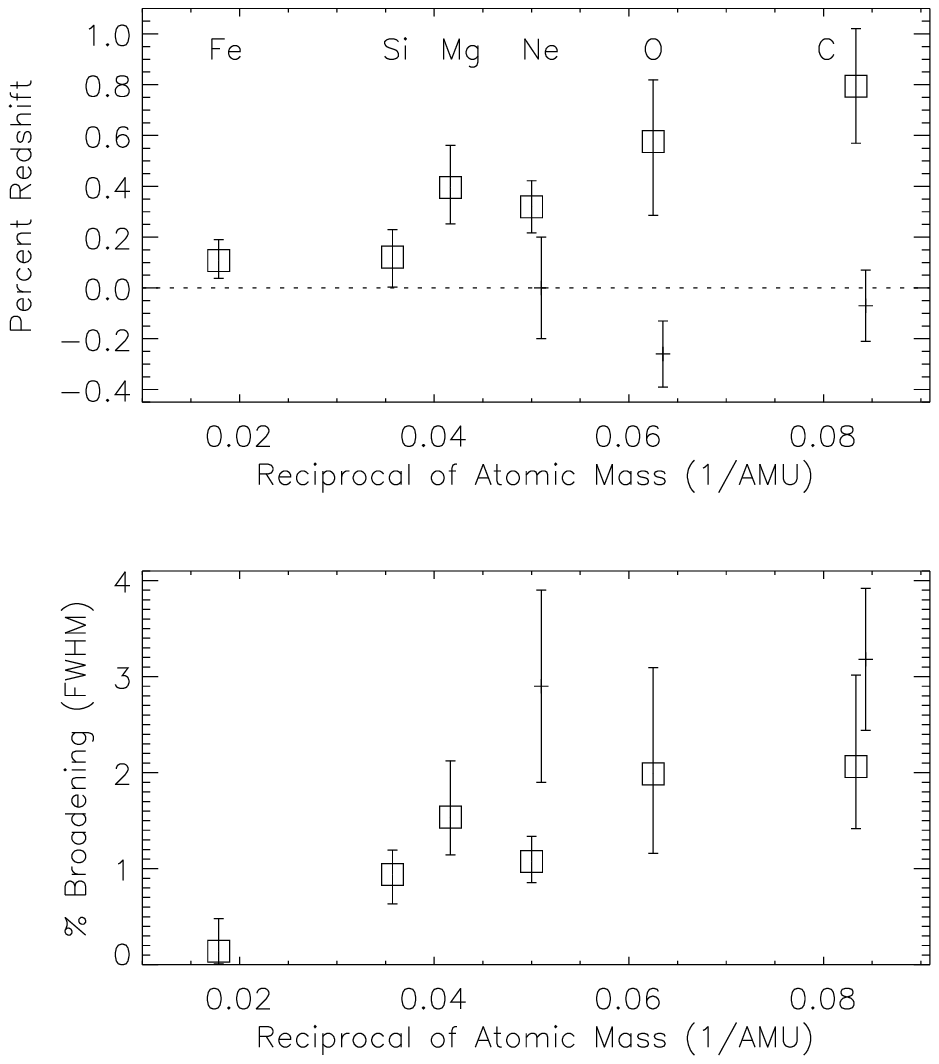}
\caption{
Redshift (top) and Doppler broadening (bottom) from Table~1
plotted as a function of the reciprocal of the mass of the nucleus.
Data points marked as squares are from this flare.  Data points
marked with dashes are the average of five flares observed by 
\it SMM/\rm GRS near $74^{\rm{o}}$ heliocentric angle.
}
\end{figure}

The full spectrum (shown in \citet{Li03}) from 250~keV to 8.5~MeV was
fitted with six bright de-excitation lines (Table~1) plus a broken
power-law representing electron bremsstrahlung (the index hardens from
2.77 to 2.23 at a break energy of 617 keV).  It was also necessary to
represent two other components of the spectrum: the very broad lines
from interactions of accelerated heavy ions with ambient hydrogen, and
the faint lines from other proton- and $\alpha$-induced lines that
cannot be measured individually.  Within the statistical limitations
of the data, the sum of these components is well-represented by three
very broad Gaussians centered at 1845, 4358 and 6575 keV.  Bright
narrow lines were included from positron annihilation \citep{Sh03a} and
neutron capture \citep{Mu03} at 511~keV and 2223~keV, respectively.
Finally, weak lines at 1263~keV and 6918~keV of intermediate width
were also included, representing expected features which are compounds
of more than one line.

The fitting was done with the Spectral Executive (SPEX) package
\citep{Sc96}, taking into account the full diagonal and off-diagonal
response of the instrument \citep{Sm02}. The spectra for the six 
bright de-excitation lines are shown in Figure~1 and the fit parameters are
given in Table~1.  The features at $\sim$840~keV in the upper left
panel of Figure~1 and $\sim$1807~keV in the lower left panel may be
the 844~keV and 1809~keV solar de-excitation lines of aluminum
\citep{Ko02}, but if so are at a surprisingly high flux.  
The effect of the instrumental resolution (about
3.0~keV at 847~keV, 4.1~keV at 1779~keV, and $\sim$ 10~keV at
6129~keV) has been removed from the width measurements, but, except
for the $^{56}$Fe line, it is negligible compared to the solar line
widths.  The best-fit Gaussians and their zero-redshift versions are
also shown in Figure~1.  Zero redshift is visibly inconsistent with
the data for most lines.

The redshifts and widths from Table~1 are shown in Figure~2.  Heavier
nuclei will recoil less from a collision with a proton or
$\alpha$-particle, and will thus show less redshift and broadening.
This trend is clearly visible.  The expected relation is
not exactly linear, since the average energy of the accelerated
particles interacting with each nucleus depends on the the cross
section for excitation.  The neon line, which appears to have a
redshift below the trend in Figure~2, has the lowest threshold for
excitation.

\section{Discussion}

The sum of the fluences in the narrow lines of Table~1 is
137~ph~cm$^{-2}$.  \citet{Sh95} calculated the total fluence in narrow
lines for 19 X-class flares observed by \it SMM/\rm GRS.  Estimating
the contribution of fainter lines not in Table~1 by comparison with
the flares studied by \citet{Sh95} gives a total line fluence of about
180 ph cm$^{-2}$ for this flare, similar to the brighter events from
\it SMM/\rm GRS, which include flares from GOES class X2.8 to X15.0.

Table~1 and Figure~2 also show \it SMM/\rm GRS results from \citet{Sh02} for the
average of five flares close to a heliocentric angle of $74^{\rm{o}}$.
The most striking difference is the lack of redshift in the lines
from carbon and oxygen, in significant disagreement with our
result for the 23 July 2002 flare.  The null redshift and high
broadening in the \it SMM/\rm GRS neon line are in marginal
disagreement with \it RHESSI \rm as well.  The advantage of high energy
resolution is clear: the \it RHESSI \rm results are more precise
using a single flare than the \it SMM/\rm GRS result for the average
of five flares, even though \it RHESSI \rm is a smaller instrument.
With more flares observed at high resolution, we will be able to tell
whether this disagreement between \it RHESSI \rm and \it SMM/\rm GRS
is due to intrinsic variations between flares or to the limitations of
\it SMM's \rm moderate-resolution spectroscopy.

\begin{figure}
\plotone{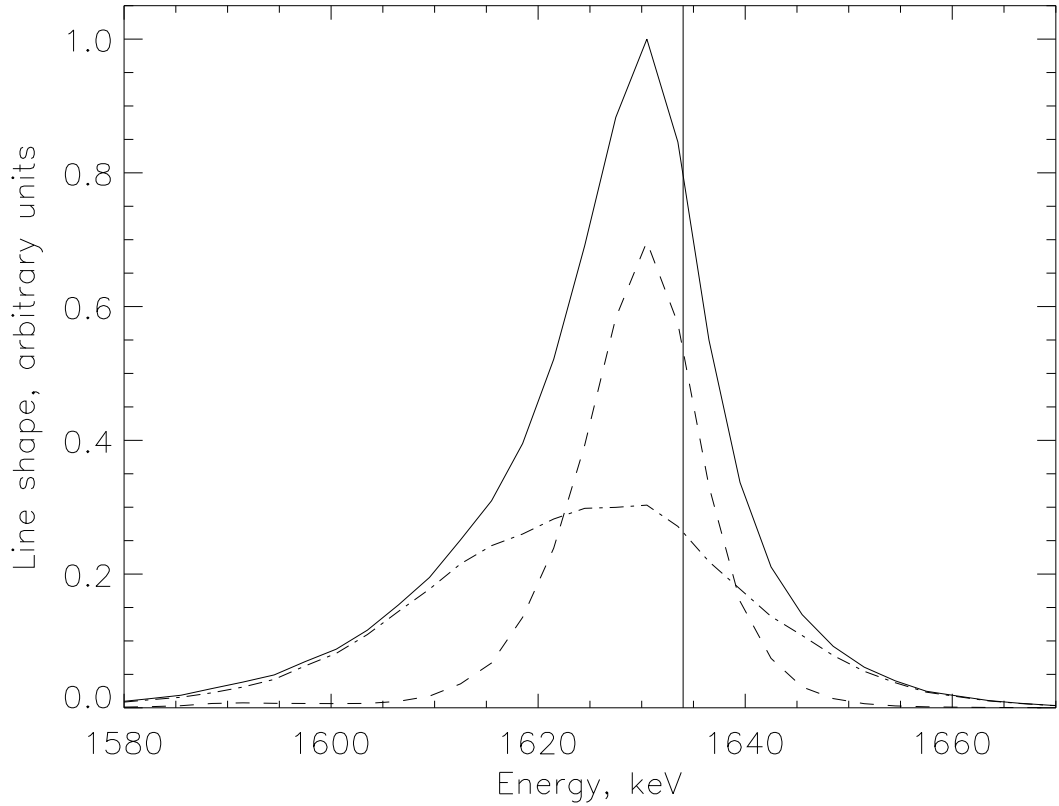}
\caption{Model $^{20}$Ne line shape for a power-law index of
--3.75, a viewing angle of 30$^{\rm{o}}$, and a 
forward-isotropic distribution.  Dashed line:
line shape from interacting protons.  Dash-dotted line:
line shape from interacting $\alpha$-particles.  Solid
curve: total line shape for an $\alpha$/proton ratio of 0.5.
Vertical line: the rest energy of the de-excitation line,
1634~keV.
}
\end{figure}

We use the modeling code of \citet{Mu88}, which allows an arbitrary
angular distribution of interacting particles about an axis at an
arbitrary angle to the observer ($\theta$).  For particles distributed
with respect to a radial field line, $\theta$ is equal to the
heliocentric angle, 73$^{\rm{o}}$ for this flare.  The simulation generates
separate gamma-ray spectra for $\alpha$-particles and protons, and
includes the effects of nuclear recoil both from the interaction and
the emission of the gamma ray.  Thus, even a purely downward beam of
incident protons gives broadened lines due to the range of recoil
angles.  

We simulated two angular distributions for the interacting ions: 1)
isotropic within the forward hemisphere and zero in the hemisphere
toward the observer, and 2) beamed directly forwards.  Since we will
explore the possibility that the magnetic field is not perpendicular
to the solar surface, we will use ``forward'' to mean along the
magnetic field away from the observer and ``downward'' to mean
perpendicular to the solar surface.

Distributions with no forward bias, such as a fan beam or a full
isotropic distribution, give no redshift and thus disagree strongly
with the redshift values in Table~1.  For each of the
distributions, the viewing angle
$\theta$ was run at 0$^{\rm{o}}$, 30$^{\rm{o}}$, 45$^{\rm{o}}$,
52$^{\rm{o}}$, 60$^{\rm{o}}$, and 73$^{\rm{o}}$.  Interpolations along
cos($\theta$) gave line shapes for intermediate angles.  The proton
and $\alpha$-particle energy spectra were modeled as power laws with
index --3.75, and the $\alpha$/proton ratio was fixed at 0.5, both
consistent with results from \it SMM/\rm GRS \citep{Sh95,Sh98}.  Figure~3 shows
the $^{20}$Ne line shape predicted by the forward-isotropic model for a
viewing angle of 30$^{\rm{o}}$.

\begin{figure}
\plotone{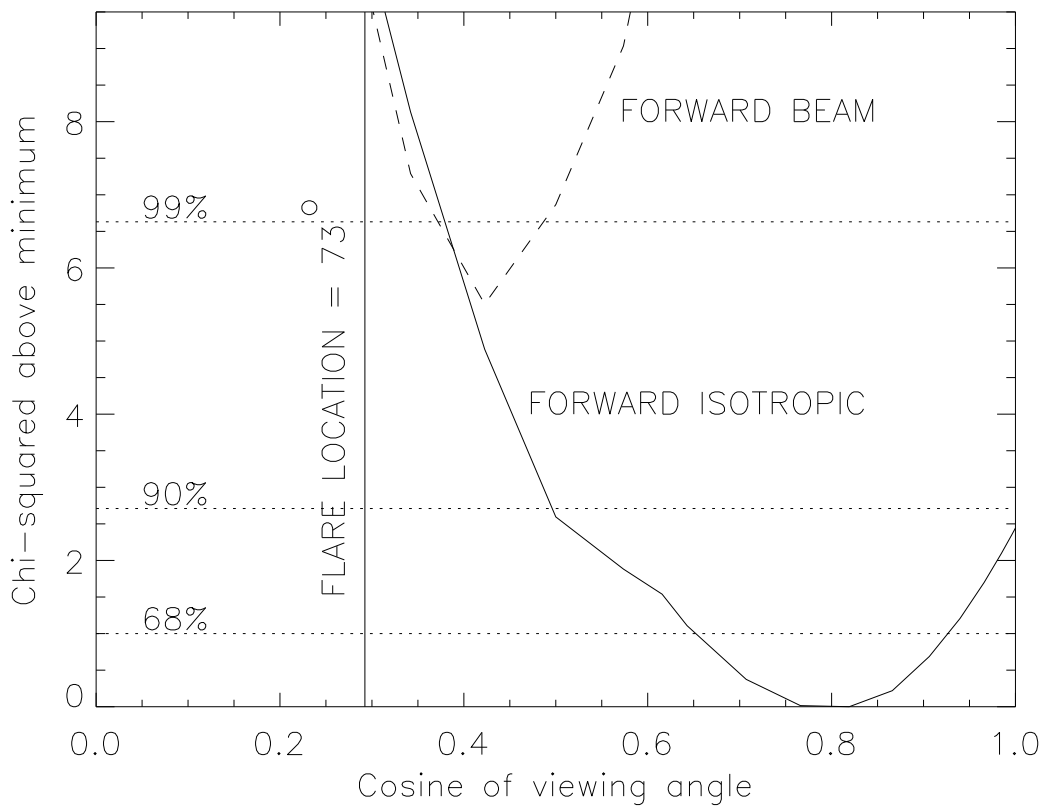}
\caption{Relative values of $\chi^{2}$ for model fits to the 23 July flare data
at various viewing angles.  Solid curve: forward-isotropic distribution.
Dashed curve: forward beam.  Dotted lines:
confidence levels for the hypothesis that the best fit is preferred
to a given model.  The left edge of the plot corresponds to a loop
viewed side-on, the right edge to a loop viewed down along the legs,
and the vertical line corresponds to a loop perpendicular to the
local solar surface.
}
\end{figure}

For both models and each viewing angle, we fitted the model lineshapes
(e.g. Figure~3) to the data, allowing the line fluences and the
underlying broad lines and continuum to vary to minimize $\chi^2$.
Figure~4 shows the relative $\chi^2$ for the two angular distributions
and all viewing angles.  The forward isotropic model is clearly a
much better fit, but it fits best with a viewing angle of
30--40$^{\rm{o}}$.  If the field lines of the loop were perpendicular
to the solar surface at their base, where the ions interact, the
viewing angle would be the heliocentric angle of the flare,
73$^{\rm{o}}$.  This
is inconsistent with the data at greater than 99\% confidence.
\it RHESSI's \rm ability to image gamma rays allows us to be certain
that the line emission comes from near the flare, 
i.e. near 73$^{\rm{o}}$ \citep{Hu03}.

We have run further simulations with power-law indices from --1.75 to
--4.75 and $\alpha$/proton ratios from
0.0 to 2.5.  In all cases the best-fit viewing angle for the
forward isotropic case is close to 30--40$^{\rm{o}}$ and
inconsistent with 73$^{\rm{o}}$.  Note that even
the forward beam, which produces the maximum possible redshift of any
distribution, still fits best with a viewing angle smaller than the
heliocentric angle.  For spectral index --2.75 and harder, the forward
beam can fit as well as the forward isotropic distribution.

A redshift could be caused either by recoil from an anisotropic
distribution of interacting ions or by bulk motion away from the
observer of the medium in which they interact.  We believe the latter
is not significant here for two reasons.  First, in the case of bulk
motion the percentage redshift is a constant, and not inversely
proportional to the mass of the excited nucleus (Figure~2).  Second,
although downward flows in the chromosphere have been observed in
impulsive flare events via red wings in H$\alpha$ line profiles, the
velocities are only 20--50~km/s \citep[e.g.][]{Wu92}, implying
redshifts of no more than 0.007-0.017\%, a trivial fraction of the
total gamma-ray shifts.

Two explanations remain for the large redshifts: either the field
lines in the loop legs are inclined toward us instead of being
perpendicular to the solar surface, or else the angular distribution
of the interacting particles is closer to a pure forward beam than a
forward-isotropic distribution.  Forward-isotropic or isotropic distributions of
interacting $\alpha$-particles are strongly favored over a downward
beam in a study of the lines at 0.429 and 0.478~MeV from
$\alpha$--$\alpha$ reactions in this flare \citep{Sh03b} and in flares
observed with \it SMM/\rm GRS \citep{Sh97}.  Also, the low average
redshifts of the $^{20}$Ne, $^{12}$C and $^{16}$O lines in all the
flares seen by \it SMM/\rm GRS \citep{Sh02}, including those used in
Table~1, favored a downward isotropic distribution (or one with a
slight additional upward component) to a downward beam.

If, on the other hand, magnetic loops containing accelerated ions are
inclined from the perpendicular for this flare, it may help explain
why the \it SMM/\rm GRS widths are larger than ours in
Table~1. \citet{Sh02} combined five flares to get their result near
$74^{\rm{o}}$, and if these had a wider range of $\theta$ than of
heliocentric angle, differing redshifts could combine to make a
broader line.  Solar active regions often do contain inclined loops,
and the magnetic topology of regions such as this one
(``$\beta\gamma\delta$'') that produce the
largest flares also tend to be the most complex.

A more detailed analysis of nuclear de-excitation line shapes
can constrain not only the angular distribution of the particles but
also the $\alpha$/proton ratio (as can be seen from
the different component shapes in Figure~3) and even the spectral
index.  The full power of the spectral analysis of gamma-ray lines,
however, will be realized when the constraints imposed by the line
shapes are combined with information from the gamma-ray line fluences
and from other observations such as gamma-ray and x-ray imaging
\citep{Hu03}, magnetograms and imaging at other wavelengths, and the
decay profile of the neutron-capture line \citep{Mu03}.

\acknowledgements

The work at the University of California, Berkeley and at NASA's
Goddard Space Flight Center was supported by NASA contract NAS5-98033,
and that at the Naval Research Laboratory by NASA DPR W19746.  We
thank Hugh Hudson, Gordon Hurford, Brian Dennis, George
Fisher, and Benzion Kozlovsky for useful discussions.


\begin{thebibliography}{}

\bibitem[Chupp et al.(1973)]{Ch73}
Chupp, E. L., Forrest, D. J., Higbie, P. R., Suri, A. N.,
Tsai, C., \& Dunphy, P. P. 1973, Nature, 241, 333

\bibitem[Hurford et al.(2003)]{Hu03}
Hurford, G. J., Schwartz, R. A., Krucker, S., Lin, R. P.,
Smith, D. M. \& Vilmer, N. 2003, \apjl, this issue

\bibitem[Kozlovsky, Murphy \& Ramaty(2002)]{Ko02}
Kozlovsky, B., Murphy, R. J., \& Ramaty, R. 2002, \apj, 141, 523

\bibitem[Lin et al.(2002)]{Li02}
Lin, R. P. et al. 2002, Solar Physics, 210, 3

\bibitem[Lin et al.(2003)]{Li03}
Lin, R. P. et al. 2003, \apjl, this issue

\bibitem[Lingenfelter \& Ramaty(1967)]{Li67}
Lingenfelter, R. E. \& Ramaty, R. 1967, in High Energy
Nuclear Reactions in Astrophysics, ed. B. S. P. Shen
(New York: W. A. Benjamin), p. 99.

\bibitem[Murphy, Kozlovsky \& Ramaty(1988)]{Mu88}
Murphy, R. J., Kozlovsky, B., \& Ramaty R. 1988, \apj, 331, 1029

\bibitem[Murphy et al.(2003)]{Mu03}
Murphy, R. J., Share, G. H., Hua, X.-M., Lin, R. P., Smith, D. M.,
\& Schwartz, R. A. 2003, \apjl, this issue

\bibitem[Ramaty and Crannell(1976)]{Ra76}
Ramaty, R. and Crannell, C. J. 1976, \apj, 203, 766

\bibitem[Schwartz(1996)]{Sc96}
Schwartz, R. A. 1996, ``Compton Gamma Ray Observatory Phase 4 Guest 
Investigator Program: Solar Flare Hard X-ray Spectroscopy,'' 
Technical Report, NASA Goddard Space Flight Center 

\bibitem[Share \& Murphy(1995)]{Sh95}
Share, G. H. \& Murphy, R. J. 1995, \apj, 452, 933

\bibitem[Share \& Murphy(1997)]{Sh97}
Share, G. H. \& Murphy, R. J. 1997, \apj, 485, 409

\bibitem[Share \& Murphy(1998)]{Sh98}
Share, G. H. \& Murphy, R. J. 1998, \apj, 508, 876

\bibitem[Share et al.(2002)]{Sh02}
Share, G. H., Murphy, R. J., Kiener, J., \& de S\'{e}r\'{e}ville, N.
2002, \apj, 573, 464

\bibitem[Share et al.(2003b)]{Sh03b}
Share, G. H., Murphy, R. J., 
Smith, D. M., Lin, R. P., Dennis, B. R., \& Schwartz, R. A. 
2003b, \apjl, this issue

\bibitem[Share et al.(2003a)]{Sh03a}
Share, G. H., Smith, D. M., Lin, R. P., Shih, A. Y.,
Hudson, H., Schwartz, R. A., Kozlovsky, B., \& Skibo, J. G. 
2003a, \apjl, this issue

\bibitem[Smith et al.(2002)]{Sm02}
Smith, D. M. et al. 2002, Solar Physics, 210, 33

\bibitem[W\"{u}lser, Canfield \& Zarro(1992)]{Wu92}
W\"{u}lser, J.-P., Canfield, R. C., \& Zarro, D. M. 1992, \apj, 384, 341

\end{thebibliography}
\end{document}